# Effect of uniform electric field on the deformation of a 2D liquid droplet in confined simple shear flow

Somnath Santra[1], Shubhadeep Mandal[2] and Suman Chakraborty[1,a]

[1]*Department of Mechanical Engineering, Indian Institute of Technology Kharagpur, Kharagpur, West Bengal - 721302, India*

[2]*Max Planck Institute for Dynamics and Self-Organization, Am Faßberg 17, D-37077 Göttingen, Germany*

In the present study, we have studied the electro-hydrodynamic of a physical system where a Newtonian dielectric liquid column or droplet suspended in another Newtonian dielectric liquid medium in presence of a simple shear flow. Taking both the phases as leaky dielectric and perfect dielectric in to consideration, we have performed 2D numerical solution for capturing the essential features of droplet deformation in between the parallel plate configuration. For a perfect dielectric system, this study shows that the deformation characteristic follows a monotonic as well as non-monotonic variation with domain confinement depending on the values of electrical permittivity ratio of the droplet and the surrounding fluid. For a leaky-dielectric system, presence of small conductivity further alters the deformation characteristic and it is happened that, at low electric field strength, the deformation increases with confinement monotonically. On contrary, deformation parameter shows non-monotonic variation with the domain confinement at higher electric field strength. Furthermore, in confined domain, the transient evolution of the deformation parameter is also markedly altered by the electric field strength in terms of steady state value of the deformation parameter and steady state time. Finally, the present analysis shows that the domain confinement significantly augment the deformation parameter in presence of electric field that leads to possible droplet break up phenomenon. From the present study, it is worthy to mention that domain confinement can be used to modulate the droplet morphology that has potential applications in modern-days droplet-based micro-fluidic devices.

## I. INTRODUCTION

The deformation and breakup of a liquid droplet and liquid column in presence of background flow under micro-confined environment is of fundamental importance in several applications. Examples are microfluidics technologies, emulsion processing, recovery of oils, atomization and spray and separation processes[1–4]. In the modern era, droplet based microfluidics become a subject of growing interest in biological and chemical field because of the advantages of low sample consumption[5], rapid mass/heat transfer[6], and high throughput[7]. Important issues in the above mentioned applications include a precious control over both the size and morphology of the droplet or liquid column during deformation. In modern day micro-fluidics devices, external electric field has been used for fine-tuned manipulation of droplet and liquid column[8–10]. For most of those cases, the droplets or liquid columns are subjected to both imposed hydrodynamic flow and electric field.

[a] E-mail address for correspondence: suman@mech.iitkgp.ernet.in

Due to the application of uniform electric field, electric stress is generated at the droplet surface. For a perfectly conducting or dielectric droplet, the electric stress acts in the perpendicular direction of the droplet surface that always tries to deform the droplet in to prolate shape [11]. But this scenario is surprisingly changed for a leaky dielectric droplet (droplet having week conductivity), where the droplet may deform into oblate or prolate shaped based on the relative magnitude of electrical conductivity, permittivity ratios between droplet and suspending media. Beside the deformation of the interface, the electric field also generates EHD flow in and around the droplet [12]. For denoting the sense of deformation, Taylor (1966) has analytically derived a deformation characteristic function ($\Omega_T$). For the case of $\Omega_T < 0$, the droplet deforms to an oblate shape, whereas for $\Omega_T > 0$, it deforms to a prolate shape. $\Omega_T = 0$ utters a zero deformation state which may occur in case of leaky dielectric due to the intricate interaction of electric and hydrodynamics normal stress. After the pioneering work of Taylor (1966)[12], several analytical, numerical and experiment studies have been made on the EHD deformation and conditions of breakup of spherical droplet[13–20]

Like EHD of droplets, in presence of electric field, the surface of the liquid column also deforms due to the generation of electrical stresses. Several studies have been made on the effect of axial and radial electric fields on the deformation characteristic of liquid columns[21–27]. But less attention is paid on the effect of transverse electric field on the EHD of liquid columns, which is of significant relevance in the continuous flow electrophoresis (CFE) technique for separating/fractioning of proteins, colloidal particles and macromolecules[28–32]. In case of CFE technique and several other cases, a 2D liquid droplet can be used as a prototypical model of liquid column. First analytical study on the effect of transverse electric field on the deformation of a liquid column (considered as 2D droplet) has been made by Rhodes et al.[28]. In the later year, some analytical and numerical analysis have been done by Esmaeeli and co-workers[33–35]. Their study shows that, the liquid column always deform into prolate shape for a perfect dielectric system, whereas for leaky dielectric system it may deform into oblate or prolate shape depending on the value of permittivity ratio and conductivity ratio of the droplet and surrounding fluid similar to EHD of droplet.

Numerous studies have also been made on the effect of a simple shear on viscous droplet suspended in a channel in the absence of electric fields [36–41]. In related studies, Sibilo et al.[42] and Mandal and Barai [43] have studied the effect of confinement in details. From their study, it is important to note that the domain confinement creates a complex oscillation in the transient deformation characteristic and droplet break up phenomenon is suppressed in the confined domain. Several studies have also been performed on the dynamics of droplet suspended under the combine action of hydrodynamic and electric flow field [44–47] where the analysis have been done considering that the droplet radius is much smaller than the channel height (electrode-electrode distance). However, the effect of domain confinement on the EHD of liquid droplet in presence of shear flow has not been explored till date. The study of micro confinement finds its relevance in several modern days micro fluidics applications, where the cross sectional area of the liquid droplet or column are typically of the same order of the domain size. Motivated by the observations, in the present study we have studied the EHD of 2D liquid droplet in confined domain in presence of simple shear flow. In the present work, our objectives are to study the effect of domain confinement on (i) steady state droplet morphology (ii) transient evolution of droplet deformation (iii) droplet break up phenomenon.

## II. PROBLEM FORMULATION

### A. System description

For the present analysis, we have considered a physical system as shown in figure. 1, where a dielectric liquid droplet is suspended in another dielectric fluid medium in confined domain. The system is subjected to simple shear flow, where the two walls of the domain are moved in the opposite direction with velocity $U$ and a uniform electric field, $E_\infty$ is also applied along the transverse direction. The physical properties of the droplet and the ambient fluid are density $\rho_i$, $\rho_e$, the viscosity, $\mu_i$, $\mu_e$, the electric permittivity, $\varepsilon_i$, $\varepsilon_e$ and the electric conductivity, $\sigma_i, \sigma_e$. The surface tension is denoted by $\gamma$. The radius of the undeformed droplet is $a$. The subscripts i and e depict the physical parameters inside and outside of the droplet respectively. Due to the application of electric field as well as imposed shear flow, the droplet deforms to a non-spherical shape and the domain confinement also effect the deformation by altering the strength of the electric field as well as viscous drag. In order to analyze the system, a two dimensional (2D) Cartesian co-ordinate has been considered and the origin is fixed at the center of the circular cross-section.

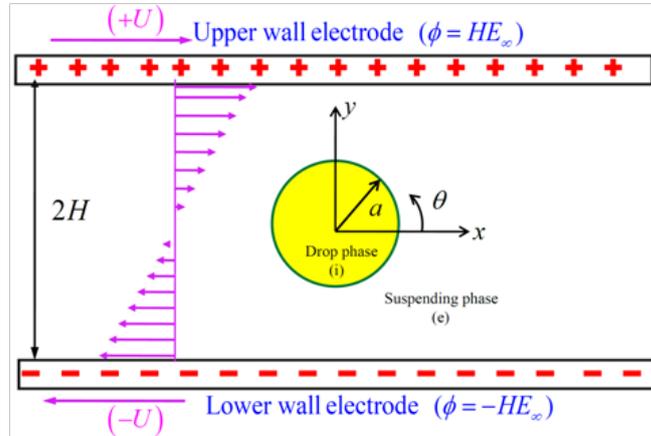

FIG.1 Schematic illustration of a physical system showing a liquid droplet of radius $a$ suspended in a confined domain encountering a uniform electric field and simple shear flow. The upper and lower wall move in the opposite direction with velocity $U$. Upper wall is connected to the positive electrode having potential $HE_\infty$ and the bottom wall is attached to the negative electrode of potential $–HE_\infty$, where $2H$ is the distance between the walls and $E_\infty$ is the magnitude of applied electric field.

### B. Governing equations and boundary conditions

In the present analysis, both the droplet and the surrounding fluids are treated as leaky dielectric fluid that have small but finite electrical conductivity. According to leaky dielectric model, the bulk fluid does not contain any free charges and free charges are present only at the interface[48]. Thus, the electric potential inside ($\phi_i$) and outside ($\phi_e$) of the droplet is represented by Laplace equation as follow

$$\left.\begin{array}{l}\nabla^2\phi_i = 0,\\ \nabla^2\phi_e = 0.\end{array}\right\} \quad (1)$$

As the electric field **E** is irrotational[33], it is related to electric potential as $\mathbf{E} = -\nabla\phi$. Inside the droplet, electric potential is bounded. Electric potential outside the droplet is specified at the two walls in the following way

$$\left.\begin{array}{l}\text{top wall: } \phi_e = HE_\infty \text{ at } y = H,\\ \text{bottom wall: } \phi_e = -HE_\infty \text{ at } y = -H,\end{array}\right\} \quad (2)$$

Here 2H is the distance between two walls. In the x-direction, electric potential is periodic in nature. In addition to this, it also fulfills the interfacial boundary conditions at the deformed interface of the droplet. At the droplet interface, electric potential is continuous

$$\phi_i = \phi_e \quad \text{at} \quad r = r_s(\theta,t), \quad (3)$$

here, $r_s(\theta,t)$ depicts the radial location of deformed interface of the droplet and $\theta$ denotes the cylindrical polar angle measured from the horizontal axis. The normal component of the current density is continuous at the droplet interface and represented in the following form

$$\mathbf{n}\cdot(R\nabla\phi_i - \nabla\phi_e) = 0 \quad \text{at} \quad r = r_s(\theta,t). \quad (4)$$

Here, $\mathbf{n} = \nabla(r - r_s)/|\nabla(r - r_s)|$ represents the outward unit normal to the droplet interface[49].

Under the paradigm of the leaky dielectric model, the fluid flow inside and outside of the droplet is governed by the continuity equation and the Navier-Stokes equation as follow

$$\begin{array}{l}\nabla\cdot\mathbf{u}_i = 0, \quad \rho\left(\dfrac{\partial\mathbf{u}_i}{\partial t} + \mathbf{u}_i\cdot\nabla\mathbf{u}_i\right) = -\nabla p_i + \nabla\cdot\left[\mu_i\left\{\nabla\mathbf{u}_i + (\nabla\mathbf{u}_i)^T\right\}\right],\\ \\ \nabla\cdot\mathbf{u}_e = 0, \quad \rho\left(\dfrac{\partial\mathbf{u}_e}{\partial t} + \mathbf{u}_e\cdot\nabla\mathbf{u}_e\right) = -\nabla p_e + \nabla\cdot\left[\mu_e\left\{\nabla\mathbf{u}_e + (\nabla\mathbf{u}_e)^T\right\}\right],\end{array} \quad (5)$$

where the velocity and pressure field are represented by **u** and *p*. For the case of sharp fluid-fluid interface limit, no volumetric electric force is involved as the bulk is charge free and also has constant permittivity. So, the governing Navier-Stokes equation also does not contain any electrical body force term and the coupling between hydrodynamic and electrostatic is done through stress balance, not through any volumetric electric force. Inside the droplet, the velocity field should remain finite and outside the droplet, the velocity field satisfies the no-slip and no penetration boundary condition at the two wall ($\mathbf{n_s}$ is unit normal vector at the solid surface) as

$$\text{top wall:} \mathbf{u} \cdot \mathbf{n}_s = 0, \mathbf{u}_e - (\mathbf{u}_e \cdot \mathbf{n}_s)\mathbf{n}_s = +U \quad \text{at} \quad y = H, \\ \text{bottom wall:} \mathbf{u} \cdot \mathbf{n}_s = 0, \mathbf{u}_e - (\mathbf{u}_e \cdot \mathbf{n}_s)\mathbf{n}_s = -U \quad \text{at} \quad y = -H, \quad (6)$$

Additionally, the velocity field also satisfies the no slip and no penetration boundary condition at the droplet interface in the following form

$$\mathbf{u}_i = \mathbf{u}_e \quad \text{at} \quad r = r_s(\theta,t), \\ \mathbf{u}_i \cdot \mathbf{n} = \mathbf{u}_e \cdot \mathbf{n} = \frac{d\mathbf{r}_s}{dt} \quad \text{at} \quad r = r_s(\theta,t). \quad (7)$$

At the interface, the balance among electric, hydrodynamic and capillary stresses that couple hydrodynamic and electrostatic in EHD flow are represented in the following form

$$(\boldsymbol{\tau}_e^H + \boldsymbol{\tau}_e^E) \cdot \mathbf{n} - (\boldsymbol{\tau}_i^H + \boldsymbol{\tau}_i^E) \cdot \mathbf{n} = \gamma(\nabla \cdot \mathbf{n})\mathbf{n} \quad \text{at} \quad r = r_s(\theta,t), \quad (8)$$

where $\boldsymbol{\tau}^H$ and $\boldsymbol{\tau}^E$ represent hydrodynamic and electric stress tensors, respectively.

### III. MATHEMATICAL CHALLENGES AND SOLUTION METHODOLOGY

First of all, for recognizing the important governing parameters, we have used the following non-dimensional scheme: length is non-dimensionalized by $a$, velocity by $S_R a$ (Where $S_R$ is the shear rate), electric field by $E_\infty$, viscous stress by $\mu_e u_s/a$, and electric stress by $\varepsilon_e E_\infty^2$. From the above non-dimensional scheme, we have fixed some non-dimensional numbers and material property ratio that are capillary number $Ca = \mu_e u_s/\gamma$ (ratio of viscous to capillary stresses), Masson number $M = \varepsilon_e E_\infty^2 a/\mu_e u_s$ (ratio of electric to viscous stresses), $S = \varepsilon_i/\varepsilon_i$, Reynolds number $Re = \rho G a^2/\mu_e$ (which signifies the relative strength of inertial stress as compared with viscous stress), viscosity ratio $\lambda = \mu_i/\mu_e$, conductivity ratio $R = \sigma_i/\sigma_e$, and permittivity ratio $S = \varepsilon_i/\varepsilon_e$. From the mathematical model, we are come to know that the deformation of the droplet interface depends on the four principle stresses: a) normal electric stress b) normal hydrodynamic stress c) viscous stress due to shear flow and d) the capillary stress at the interface. It is worthy to mention that the droplet shape is not known from priori and the calculation of it needs the idea of electric stress and hydrodynamic stress at the droplet interface. Again, the flow field and electric potential associated with the electric stress and hydrodynamic stress depends on the knowledge of the droplet shape that creates a non-linear boundary condition at the droplet interface even underneath the creeping flow ($Re=0$) condition. Due to that reason, we cannot get the analytical solution for the arbitrary values of non-dimensional number. Furthermore, another complication comes from the confined parallel plate configuration. Because of that, we have adopted the numerical solution of the problem for capturing the dynamics of droplet interface in presence of transverse electric field and simple shear flow in confined domain. Neglecting the effect of wall confinement, we have also performed an analytical solution and compared them with the numerical results for checking the correctness of numerical result (refer to appendix A for analytical solution).

In the present analysis, Phase-field method [50,51] has been used for representing the fluid-fluid interface for highly deformed droplet. The main reason behind the use of Phase-field method is that it has no necessity of tracking the interface as the diffuse interface takes the place of sharp interface. In the present days, this formalism have been used with good accuracy and reported in several studies [52–54]. In the phase field method, an order parameter $\varphi$ (x, t) has been used for the determination of the distribution of constituting fluid element. According to figure 1, $\varphi$ (x, t) =1 in the suspending fluid region whereas $\varphi$ (x, t) =- 1 in the inner region of the droplet. The interfacial region has been denoted by a diffuse zone expressed as $-1 < \varphi < 1$. For the dynamic evolution of the order parameter $\varphi$, Cahn-Hilliard equation has been used. The equation is represented in the following dimensional form

$$\frac{\partial \varphi}{\partial t} + \mathbf{u} \cdot \nabla \varphi = \nabla \cdot \left( M_\varphi \nabla G \right), \quad (9)$$

where $M_\varphi$ is the interface mobility factor that determines the relaxation time of the interface and time scale of the Cahn–Hilliard diffusion. $G = \gamma \left( \varphi^3 - \varphi \right) / \zeta - \gamma \zeta \nabla^2 \varphi$ is the chemical potential. Here $\zeta$ is a parameter which controls the thickness of diffuse interface. In the non-dimensional form, Eq. (9) can be expressed as

$$\frac{\partial \overline{\varphi}}{\partial t} + \overline{\mathbf{u}} \cdot \overline{\nabla} \overline{\varphi} = \frac{1}{Pe} \overline{\nabla} \cdot \overline{M}_f \left( \overline{\nabla} \overline{G} \right); \text{ where } \overline{G} = \frac{1}{Cn} \left( \overline{\varphi}^3 - \overline{\varphi} \right) - Cn \overline{\nabla}^2 \overline{\varphi} \quad (10)$$

Electric potential can be obtained by solving following governing equation,

$$\overline{\nabla} \cdot (\overline{\sigma} \overline{\nabla} \overline{\phi}) = 0 \quad (11)$$

The pressure and velocityfield can be obtained by solving the continuity equation and Cahn-Hilliard Navier-Stokes equation that couple the phase field formalism with the EHD, expressed as follow

$$\overline{\nabla} \cdot \overline{\mathbf{u}} = 0 \quad (12)$$

$$Re \left( \frac{\partial \overline{\mathbf{u}}}{\partial t} + \overline{\nabla} \cdot (\overline{\mathbf{u}}.\overline{\mathbf{u}}) \right) = -\overline{\nabla} \overline{p} + \overline{\nabla} \cdot \left[ \left\{ \overline{\nabla} \overline{\mathbf{u}} + (\overline{\nabla} \overline{\mathbf{u}})^T \right\} \right] + \frac{1}{Ca} \overline{G} \overline{\nabla} \overline{\varphi} + M \overline{\mathbf{F}}^E. \quad (13)$$

where, $\overline{\mathbf{F}}^E = \overline{\nabla} \cdot \left( \overline{\varepsilon} \overline{\nabla} \overline{\phi} \right) \overline{\nabla} \overline{\phi} - \left| \overline{\nabla} \overline{\phi} \right|^2 \overline{\nabla} \overline{\varepsilon} / 2$ and $\overline{G} \overline{\nabla} \overline{\varphi}$ denotes the the phase field dependent surface force because of interfacial tension. In the paradigm of the phase field formalism, the fluid properties are defined in the following way

$$\left.\begin{aligned}\rho &= \frac{(1+\varphi)}{2}\rho_r + \frac{(1-\varphi)}{2} \\ \mu &= \frac{(1+\varphi)}{2}\lambda + \frac{(1-\varphi)}{2} \\ \varepsilon &= \frac{(1+\varphi)}{2}S + \frac{(1-\varphi)}{2} \\ \sigma &= \frac{(1+\varphi)}{2}R + \frac{(1-\varphi)}{2}\end{aligned}\right\} \quad (14)$$

In Eq.(10)-(13), the involved non-dimensional parameters are: Péclet number $Pe = (a^2 u_s / M_\varphi \gamma)$, Reynolds number, $Re = \rho u_s a / \mu_0$, capillary number $Ca = \mu u_s / \gamma$ and Cahn number $Cn = \zeta / a$. $Ca_E = \varepsilon_o E_\infty^2 a / \gamma$ represents the electric capillary number and finally, Masson number is shown by $M = Ca_E / Ca$. One must acknowledge that periodic boundary condition has been used for velocity and pressure fields as well as phase field variable in the horizontal direction in the following shape,

$$\left.\begin{aligned}&(i) \quad \mathbf{u}(\mathbf{x}) = \mathbf{u}(\mathbf{x}+L), \\ &(ii) \quad p(\mathbf{x}) = p(\mathbf{x}+L), \\ &(iii) \quad \varphi(\mathbf{x}) = \varphi(\mathbf{x}+L).\end{aligned}\right\} \quad (15)$$

.For solving the governing equations (Eq.(10)-(13)), we have used finite element solver COMSOL Multiphysics.

## IV. RESULTS AND DISCUSSIONS

### A. Validation of the numerical results

First of all, we have compared the numerical result with the analytical solution. The analytical solution is developed in a unbounded domain by using regular perturbation method taking $Ca$ as a perturbation parameter (shown in appendix A). In regular perturbation method, the expansion of the droplet radius in asymptotic form can be represented in the following shape [55,56],

$$r_s = 1 + f(\theta) = 1 + Ca f^{(Ca)} + Ca^2 f^{(Ca^2)} + O(Ca^3) \quad (16)$$

In Eq.(16), $f^{(Ca)}$ and $f^{(Ca^2)}$ show the deviation in droplet shape under deformed condition and expressed as,

$$f^{(Ca)} = \left( L_{2,0}^{(Ca)} \cos(2\theta) + \hat{L}_{2,0}^{(Ca)} \sin(2\theta) \right)$$

$$f^{(Ca^2)} = L_{0,1}^{(Ca^2)} + \begin{bmatrix} L_{2,1}^{(Ca^2)} \cos(2\theta) + \hat{L}_{2,1}^{(Ca^2)} \sin(2\theta) + L_{3,1}^{(Ca^2)} \cos(3\theta) \\ + \hat{L}_{3,1}^{(Ca^2)} \sin(3\theta) + L_{4,1}^{(Ca^2)} \cos(4\theta) + \hat{L}_{4,1}^{(Ca^2)} \sin(4\theta) \end{bmatrix} \quad (17)$$

The expression of $L_{2,0}^{(Ca)}$ and $\hat{L}_{2,0}^{(Ca)}$ have been obtained as

$$L_{2,0}^{(Ca)} = -\frac{1}{3} M \frac{\Omega_T}{(R+1)^2} \left( 1 - e^{-\frac{t}{\lambda+1}} \right), \quad \hat{L}_{2,0}^{(Ca)} = \left( 1 - e^{-\frac{t}{\lambda+1}} \right). \quad (18)$$

The sense of interface deformation depends on $\Omega_T$ (deformation characteristic function), expressed as $\Omega_T = R^2 + R - 3S + 1$ [48]. $\Omega_T > 0$ denotes that the droplet will deform into prolate shape (major axis of ellipse is directed along the direction of electric field) where as $\Omega_T < 0$ necessarily means that the droplet will deform into oblate shape (major axis of ellipse is directed perpendicular to the direction of electric field). Different coefficients present in the higher-order shape correction are given in appendix A. The degree of droplet deformation has been measured by droplet deformation parameter $D$ and expressed as

$$D = \frac{\max r_s(\theta) - \min r_s(\theta)}{\max r_s(\theta) + \min r_s(\theta)}. \quad (19)$$

Here, $\theta$ is the droplet orientation angle that is formed by the major axis of elliptical droplet with the direction of shear flow.

From now onwards, the effect of different governing parameters on $D$ has been discussed. For that purpose, we have chosen three perfect dielectric system that are system I, II, and III having $S=15$, $S=2$ and $S=0.1$ respectively. Furthermore, we have also considered four leaky dielectric systems having $(R, S)=(2, 0.5)$, $(R, S)=(10, 1)$, $(R, S)=(0.033, 0.4397)$ and $(R, S)=(0.5, 2)$ which are defined as system IV, V, VI and VII respectively. For checking the accuracy of the present numerical simulation, we have made a comparison between the analytical result and the numerical result on steady state deformation parameter $(D_\infty)$, that is obtained analytically in the limit $t \to \infty$. For getting steady state values in numerical simulation, we have allowed long time to run the simulation. For unbounded consideration, we have chosen $Wc=0.2$. Figure 2(a) shows the variation of $D_\infty$ with $Ca$ for a perfect dielectric system I, whereas Fig. 2(b) and 2(c) show the variation of $D_\infty$ with $Ca$ for the system IV and system VII, respectively. From the three figures, it is cleared that both the leading-order linear theory corrected up to $O(Ca)$ and the higher-order nonlinear theory corrected up to $O(Ca^2)$ agree very well with the numerical solution for lower values of $Ca$. But, the higher-order nonlinear theory shows better matching with the numerical solution with respect to the leading-order linear theory for comparatively higher value of $Ca$.

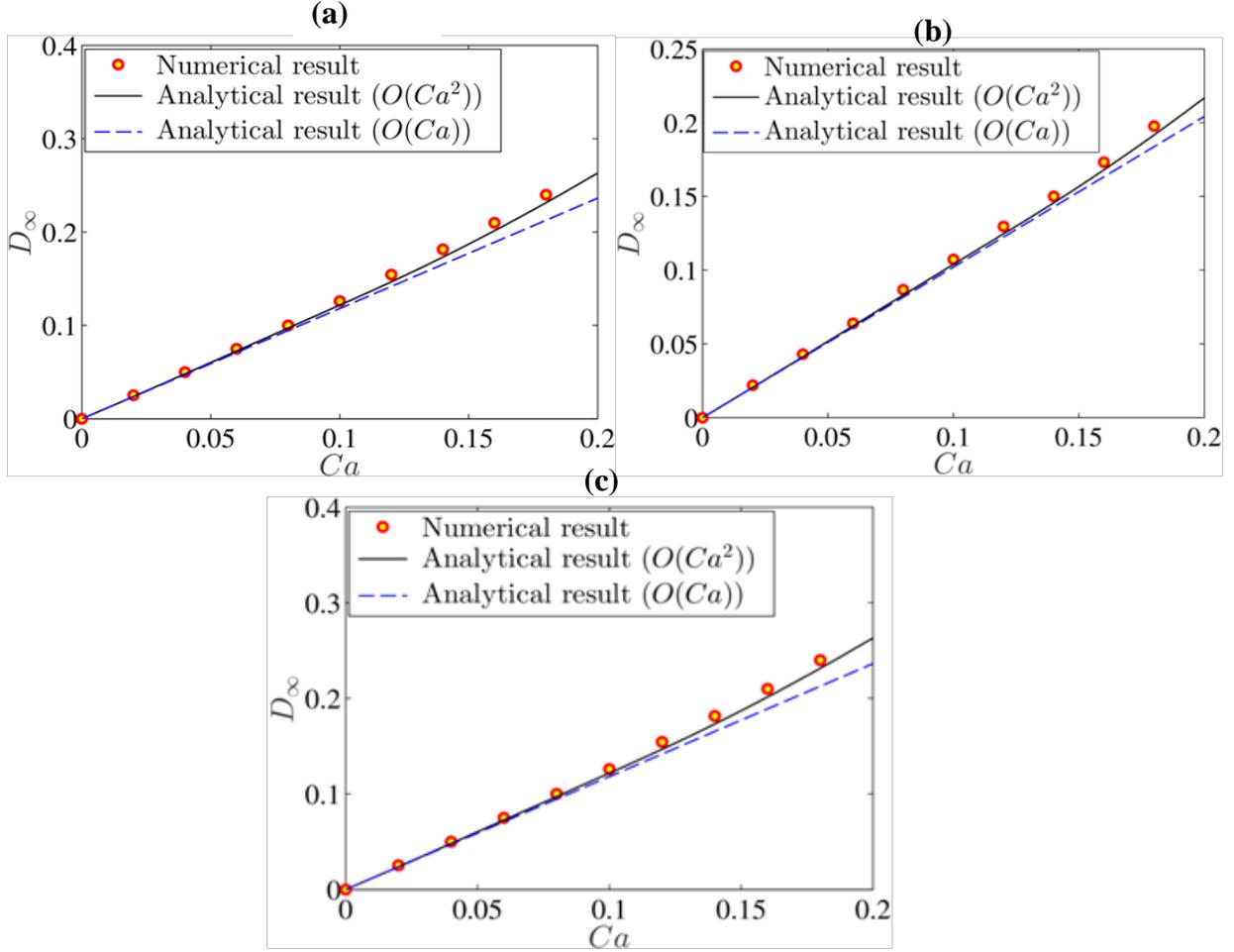

FIG.2 Variation of $D$ with $Ca$ for (a) perfect dielectric system I with $S = 15$ (b) leaky dielectric system IV with $(R, S)$=(2, 0.5) (c) leaky dielectric system VII with $(R, S)$=(0.5, 2). Others parameters are $M$ =1, $\lambda$=1, $Wc$=0.2, $Re$ = 0.01

## B. Effect of domain confinement on steady state deformation

Figure 3 shows the effect of domain confinement on steady state deformation parameter ($D_\infty$). We have varied confinement ratio ($Wc$) from 0.2 (unbounded domain) to 0.8 (extremely confined domain). Figure 3(a) depicts the variation of steady state deformation parameter with $Wc$ for the perfect dielectric system II. From the figure, it has been obtained that the value of $D_\infty$ decreases with $Wc$ upto $Wc \approx 0.30$ and beyond it, further increase in the domain confinement results in the enhancement of $D_\infty$ both at lower and higher electric field strength ($M$=1 and $M$=10). A similar phenomenon is also observed for perfect dielectric system III at higher electric field strength ($M$=10) where the magnitude of $D_\infty$ decreases with confinement ratio up to $Wc$≈0.5 and beyond it, if we further increase the value of $Wc$, the steady state deformation again increases as shown in Fig 3(b). On contrary to this, $D_\infty$ increases monotonically with domain confinement at lower electric field strength ($M$=1). This droplet deformation characteristic is occurred due to the interplay between normal electric stress (as normal hydrodynamic stress is not present), viscous stress and the effect of domain confinement on the stress. It is important to mention that the strength of the normal electric stress depends on electric field strength as well as domain confinement, where as viscous stress is only governed by the domain confinement for a given shear flow [42,43]. The normal electric stress always tries to deform the droplet into prolate shape where as viscous stress tries to deform it along the direction of flow and retard the effect of normal electric stress. For perfect dielectric system II (for considered value of $S$), increase of viscous stress with confinement is comparatively higher than the normal electric stress as the value of $S$ is not much high. In unbounded domain, the normal electric stress is dominating as the strength of shear stress is low. But with increase in $Wc$, the magnitude of shear stress increases that retard the deformation via reducing the effect of normal electric stress and the deformation achieves its minimum value at $Wc=0.3$. After $Wc=0.3$, the viscous stress becomes dominating and further increase in $Wc$ again enhances the deformation due to increase in shear stress and the droplet aligns more in the flow direction. A similar interplay between normal electric stress and viscous stress is also responsible for the observed deformation characteristic of perfect dielectric system III at higher electric field strength. On the other side, at lower electric field strength ($M=1$), the deformation increases monotonically with increases in $Wc$ due to the dominating nature of shear stress over the normal electric stress.

Similarly, Fig. 3(c) shows the variation of $D_\infty$ with $Wc$ for system IV for different values of Masson number ($M=1$ and $M=20$). From Fig. 3(c), it is cleared that the deformation of the droplet is higher in low confined zone and with increase in the domain confinement, the value of steady state deformation decreases for $M=20$, where as for $M=1$, the values of $D_\infty$ increases monotonically with $Wc$. The physical explanation of the above observed behavior is now provided. For the present case, the normal hydrodynamic stress also acts in addition to normal electric stress and viscous drag stress. It is worthy to mention that the normal hydrodynamic stress decreases with increase in $Wc$ and above a critical domain size, it becomes reversed. If we further increase the value $Wc$, its magnitude again increases [35]. For the considered values of ($R$, $S$), at lower domain confinement, both the normal electric and hydrodynamic stress act in the same direction at lower domain confinement and tries to deform the droplet into prolate shape, whereas viscous drag stress tries to deform it along the flow direction though its magnitude is small due to low confinement. For the case of higher electric field strength ($M=20$), the combine strength of normal electric and hydrodynamic stress is more that creates higher deformation in low confined domain. With increase in the confinement ratio, both normal electric stress and viscous drag stress increase, whereas the magnitude of normal hydrodynamic stress decreases and after certain confinement ratio (here, $Wc=0.5$), it becomes reverse. So, in confined domain, the effect of normal electric stress is reduced by the combine effect of viscous stress and reversed normal hydrodynamic stress. Due to that reason, the deformation decrease with increase in $Wc$ and the droplet tries to align more in the flow direction. For lower electric field strength ($M$=1), the droplet deformation characteristic is regulated by the enhancement of viscous drag with domain confinement. So, it is monotonically increases.

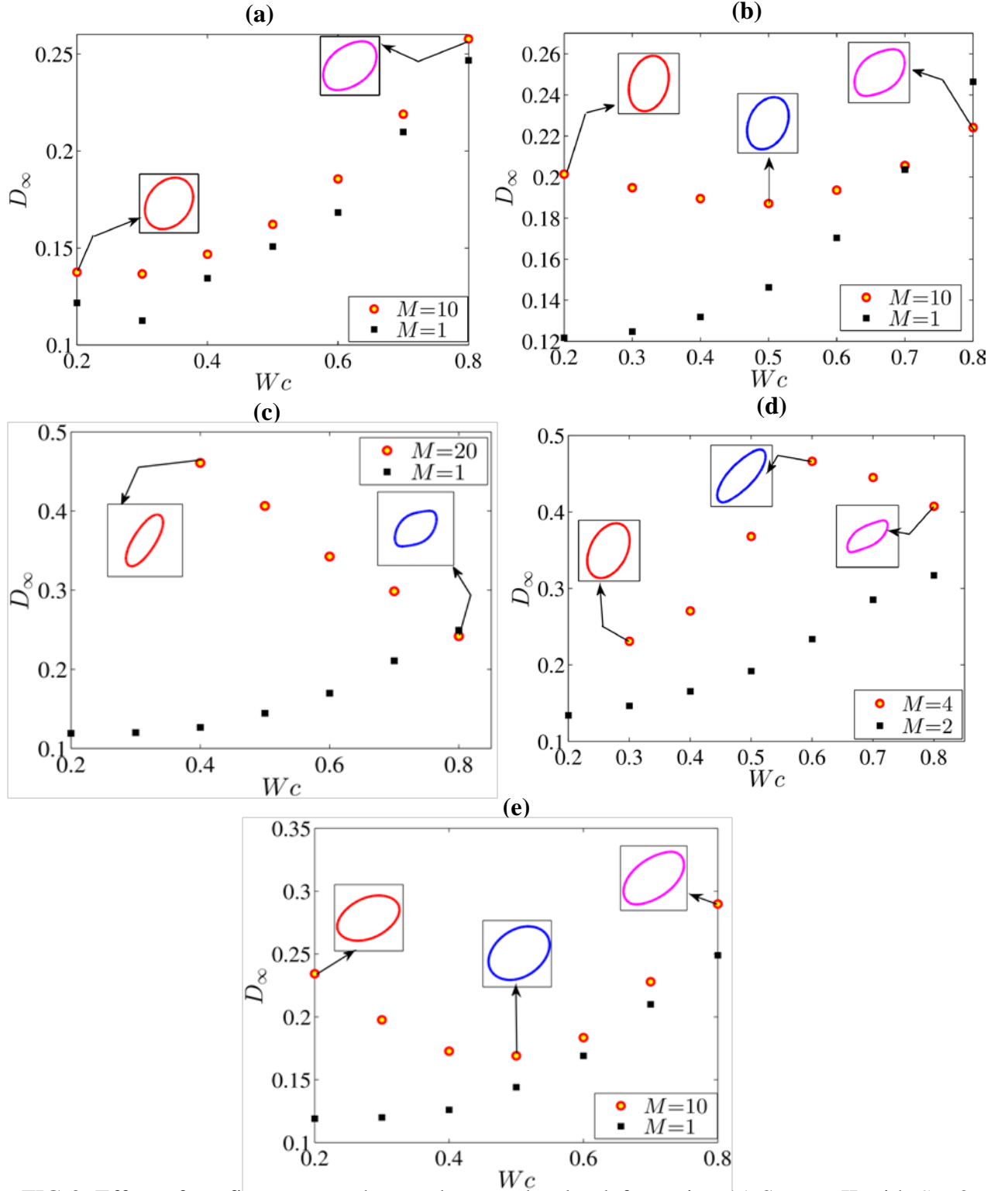

FIG 3. Effect of confinement on the steady-state droplet deformation (a) System II with $S = 2$ (b) system III with $S = 0.1$ (c) system IV with $(R, S)=(2, 0.5)$ (d) system V with $(R, S)=(10, 1)$ (e) system VI with $(R, S)=(0.033, 0.4397)$. Others parameters are $Ca=0.1$, $\lambda=1$, $Wc=0.2$

Figure 3(d) shows the variation of steady state deformation parameter with confinement ratio for leaky dielectric system V. From the figure, it has been obtained that the steady state deformation parameter increases with increase in the $Wc$ for lower electric field strength. But, at comparatively higher electric field strength, the deformation increases with $Wc$ up to $Wc$=0.8 and further increase in $Wc$ reduces the deformation. The physical interpretation of the above observed deformation characteristic as shown in Fig. 3(d) is now provided. For the present case also, droplet deformation characteristic is governed by the normal electric stress, normal hydrodynamic stress, viscous drag and the effect of domain confinement on the stresses. For the considered value of ($R$, $S$), normal electric stress and hydro dynamic stress deform the droplet toward the electrode in unbounded domain at higher value of electric field strength ($M$=4). On contrary, the viscous drag tries to deform it toward the flow direction. With increase in the $Wc$, the strength of normal electric stress and viscous drag increases, whereas hydrodynamic normal stress decreases. For the present case, as the liquid droplet is more conducting than the ambient fluid ($R$=10), the effect of domain confinement on normal electric stress is pronounced. So, the deformation increases with $Wc$ upto $Wc$ =0.8. But after $Wc$=0.8, the combined effect of reversed normal hydrodynamic stress and viscous drag becomes dominating that reduces the deformation by decreasing the effect of normal electric stress and it aligns the droplet more toward the flow direction. The droplet deformation phenomenon observed at lower electric field strength can be explained in the similar way like system IV.

Figure 3(e) shows the variation of deformation parameter with confinement ratio for leaky dielectric system VI. From the figure, it is cleared that the steady state deformation parameter increases with confinement ratio at lower electric field strength ($M$=1) like other leaky dielectric system. But the deformation scenario gets surprisingly changed when the strength of the electric field increases, where the steady state deformation value decreases with $Wc$ upto $Wc$≈0.5 and further increase in $Wc$ again increases the deformation. The reason behind the observed droplet deformation phenomenon is now presented. For the considered values of ($R$, $S$), the normal electric stress and normal hydrodynamic stress try to deform the droplet along the flow direction similar to viscous drag in unbounded domain. At higher electric field strength, with increase in the domain confinement, the normal hydrodynamic stress reduces significantly that reduces the deformation. After $Wc$≈0.5, the normal hydrodynamic stress become reversed and if we further increase $Wc$, the magnitude of reversed normal hydrodynamic stress increases. In confined domain, the higher magnitude of reversed hydrodynamic stress creates higher deformation and tries to align the droplet along the electrode.

## C. Effect of domain confinement on transient droplet deformation characteristics

Figure 4 depicts the transient evolution of the deformation parameter for different value of $Ca_E$. We have shown both the analytical result obtained from higher-order nonlinear theory corrected upto $O(Ca_E^2)$ (that is shown by line) and numerical result (shown by markers) and performed a comparison between them. The variation of $D$ with time for a perfect dielectric system I has been shown in Fig. 4(a), where as Fig. 4(b) and 4(c) show the variation for system IV and system VII respectively. From all the figures, it is cleared that the analytical result shows a well agreement with the numerical result for lower value of $Ca_E$. But, at higher values of $Ca_E$, the deviation is seen to be higher where the analytical result under predicts the deformation characteristic for all the systems.

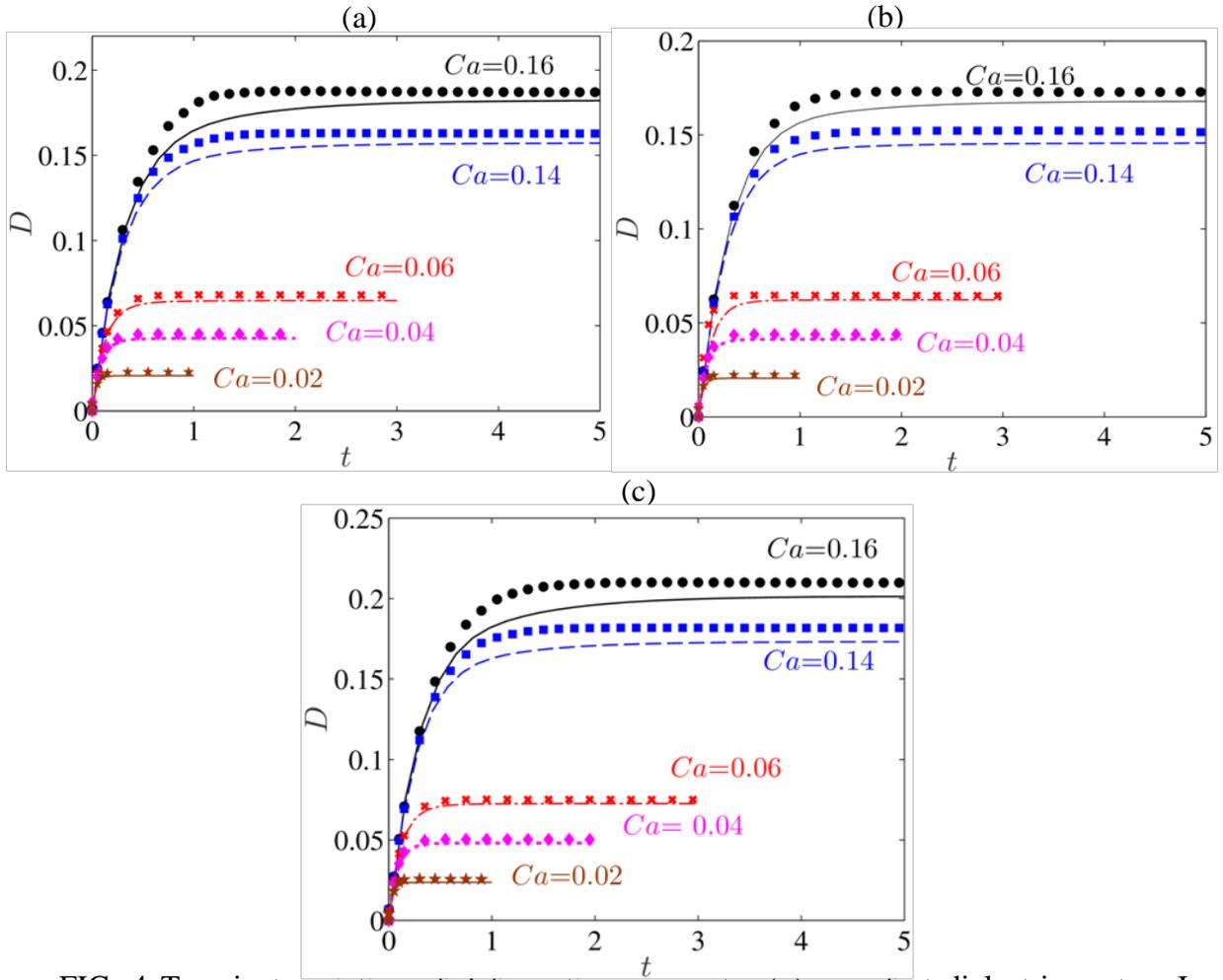

FIG. 4 Transient variation of deformation parameter (a) a perfect dielectric system I with $S$ =15, (b) system IV with $(R, S)$=(2, 0.5) (c) system VII with $(R, S)$=(0.5, 2). Others parameters are $M$=1, $\lambda$=1, $Wc$ =0.2, $Re = 0.01$

For analyzing the transient droplet deformation characteristic in confined domain under different electric field strength, we have plotted Fig. 5, Fig. 6, Fig.7 and Fig. 8. In this section, we have varied the values $M$ via changing the values of $Ca_E$ while keeping the value of $Ca$ is fixed. Figure 5(a) and 5(b) show the variation of $L$ and $\theta$ with time respectively for a perfect dielectric system II with $S=2$ for different values of $M$. Figure 5(a) shows that with increase in the value of $M$, the oscillation in the deformation characteristic increases that leads to higher value of steady state time. On contrary, the steady state value of deformation parameter reduces ($L_\infty = 1.992$ for $M=5$ and $L_\infty = 2.07$ for $M=0$).

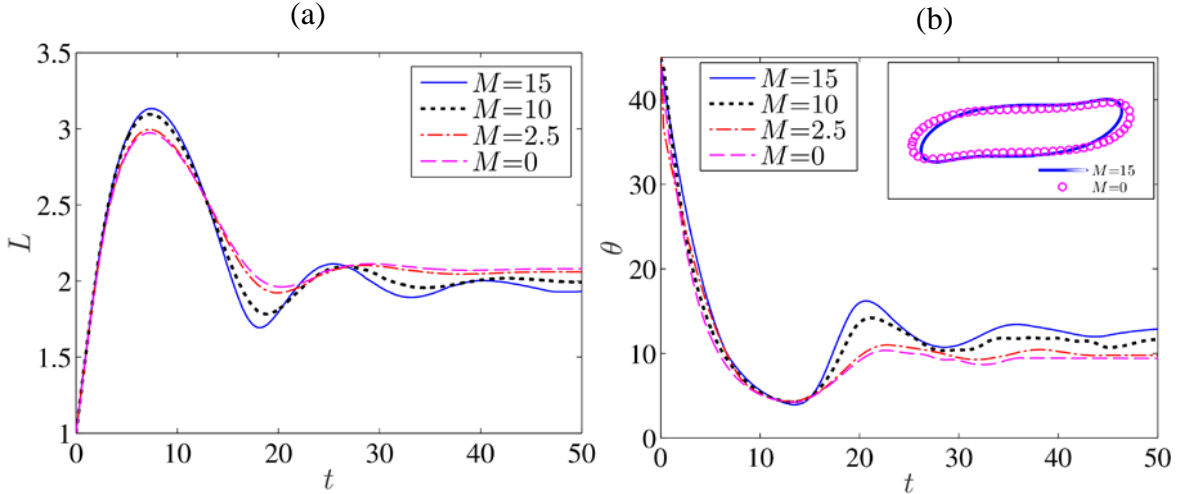

FIG. 5. Variation of (a) $L$ and (b) $\theta$ with $t$ for a perfect dielectric system II with $S=2$. Other parameters are $Ca=0.4$, $\lambda=1$, $Re=0.01$, $Wc=0.909$. The inset shows the contours of the droplet at steady state for $M=0$ and $M=15$

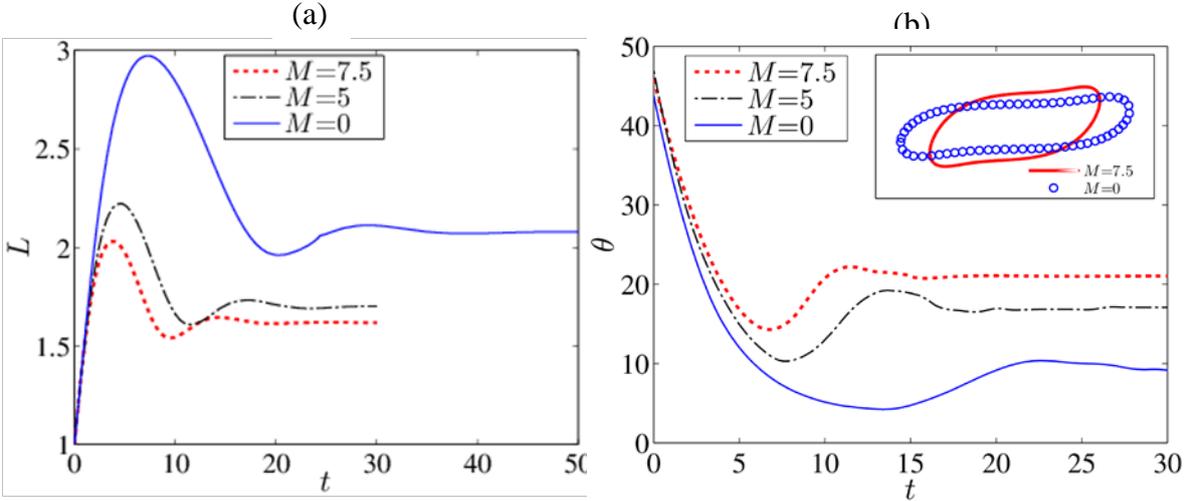

FIG. 6. Variation of (a) $L$ and (b) $\theta$ with $t$ for a perfect dielectric system III with $S=0.1$. Other parameters are $Ca=0.4$, $\lambda=1$, $Re=0.01$, $Wc=0.909$. The inset shows the contour of the droplet at steady state for $M=0$ and $M=7.5$

Figure 6(a) and 6(b) depict the variation of $L$ and $\theta$ with time respectively for a perfect dielectric system III. From Fig. 6(a), it is noted that the value of steady state deformation parameter decrease with increase in the value of $M$ ($L_\infty=1.617$ for $M=7.5$ and $L_\infty=2.07$ for $M=0$) like perfect dielectric system II. But the interesting thing is that the value of steady state time also decreases with increase in the values of $M$ unlike the perfect dielectric system II.

For both the perfect dielectric system, it is well established that the deformation of droplet interface is governed by the normal electric stress due to electric field and the viscous drag due to the shear flow. For a higher value of *Ca* (here *Ca*=0.4) and *Wc*, the shear stress generates a large viscous drag and tries to deform the droplet along its direction. On the other side, the electric normal stress tries to deform it in prolate shape that is perpendicular to the shear plane. For both the system, it is happened that at lower value of *M*, the viscous drag is higher than the normal electric stress in confined domain due to higher values of *Ca* and regulates the droplet deformation phenomenon. As we increase the value *M* (or $Ca_E$), the normal electric stress increases that reduces the deformation via lowering the effect of viscous drag and tries to deform the droplet along its direction. So, the droplets are aligned more in the direction of electrode as shown in Fig. 5(b) and Fig. 6(b). The unsteadiness in the deformation behavior is occurred due to continuous elongation and relaxation of the droplet interface. It is necessary to say that this elongation and relaxation is coupled with forward and backward rotation of the droplet that is occurred at higher shear rate and the interplay among the normal electric force, viscous drag force and surface tension force finally determine the time at which the droplet achieve steady state configuration.

Figure 7(a) and 7(b) depict the transient evolution of *L* and *θ* respectively for leaky dielectric system VII. From Fig.7(a), it is cleared that the unsteadiness in the transient deformation characteristic increases with *M* that creates a higher steady state time. Likely, the steady state value of *L* also increases with *M* ($L_\infty$=2.07 for *M*=0 and $L_\infty$=2.301, *M*=0.75).

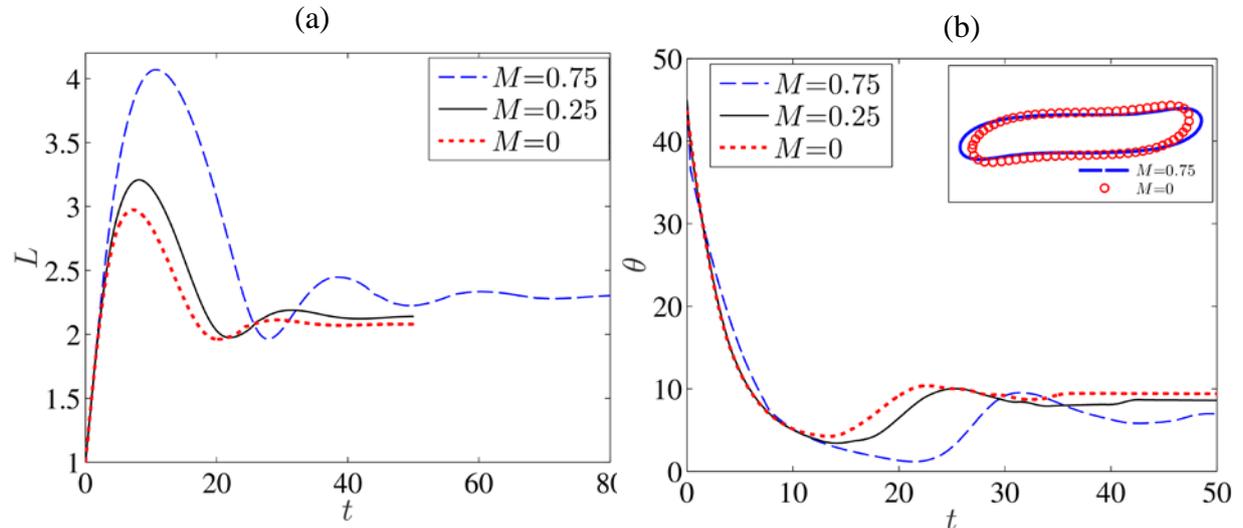

FIG. 7. Variation of (a) *L* and (b) *θ* with *t* for system VII with (*R*, *S*)=(0.5, 2). Other parameters are *Ca*=0.4, *λ*=1, *Re*=0.01, *Wc*=0.909. The inset shows the contour of the droplet at steady state for *M*=0 and *M*=0.75

For explaining this deformation dynamics, the role of four stresses that are a) normal electric stress b) normal hydrodynamic stress c) viscous drag and d) capillary stress have been taken into consideration. For the present case, due to having high value of *Ca* (*Ca*=0.4), the viscous force is very much stronger in highly confined domain (*Wc*=0.909) and contributes much in regulating the deformation. Furthermore the normal electric is also high in confined domain[35]. For the considered values of (R, S), the combine action of viscous stress and normal electric stress tries to deform the droplet into oblate shape (along the flow direction) whereas the hydrodynamic normal stress tries to deform it into prolate shape. With increase in the value of M, combined magnitude of viscous stress and electric normal stress has been increased that ultimately increases the value of L and tries to align the droplet along the flow direction as shown in figure 7(b). For the present case as deformation increases with M, the interface deforms more for locally balance the stresses that increases the time to achieve steady state configuration.

Similarly, Fig. 8(a) and 8(b) also illustrate the transient variation of L and $\theta$ respectively for different values of M for leaky dielectric system IV. But unlike system VII, for the present case, both the steady state time and steady state value of deformation parameter decrease with increase in M ($L_\infty$=1.749 for M=7.5 and $L_\infty$=2.07 for M=0) as shown in Fig. 8(a).

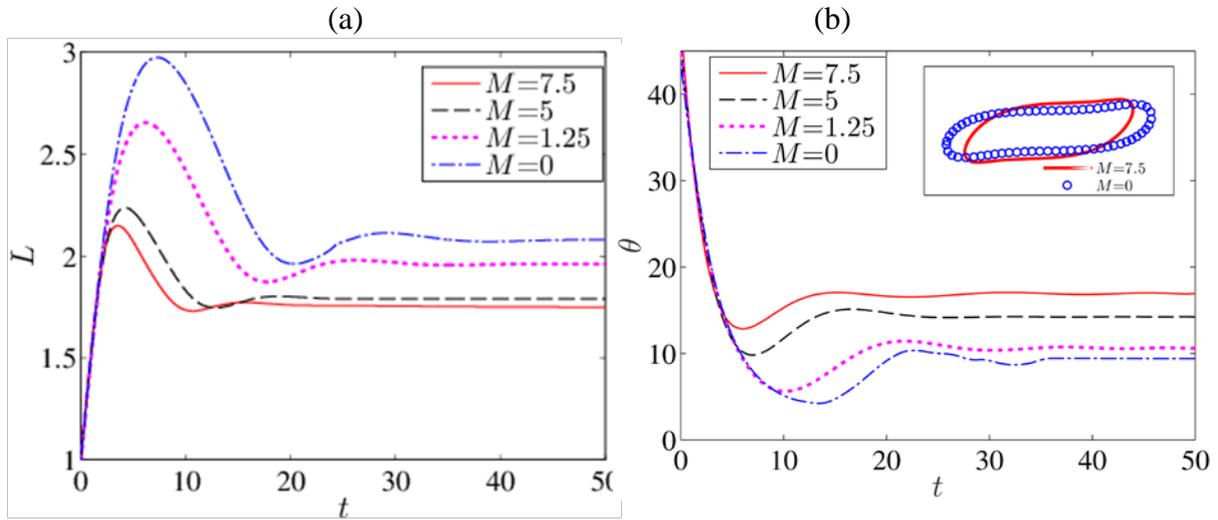

FIG. 8. Variation of (a) L and (b) $\theta$ with t for system IV with (R, S)=(2, 0.5). Other parameters are Ca=0.4, $\lambda$=1, Re=0.01, Wc=0.909. The inset shows the contour of the droplet at steady state for M=0 and M=7.5

For the considered value of permittivity and conductivity ratio (R≫S), the magnitude of normal electric stress (as it is quadratic in R) is much higher than the normal hydrodynamic stress (it acts in opposite to normal electric stress) and net normal stress tries to deform the droplet along the vertical direction (prolate shape), whereas the viscous stress tries to deform it along the direction of flow. In highly confined domain, the higher viscous stress creates a large droplet deformation along the flow direction at higher value of Ca. As M increases, the net normal stress also increases that lowers the deformation parameter via reducing the effect of viscous drag and tries to align the droplet along vertical direction as shown in figure 8(b). For that reason, the value of steady state deformation decreases with increase in M and steady state configuration has been reached at comparatively faster rate.

## D. Effect of domain confinement on droplet break up

Figure 9(a) and 9(b) show the influence of domain confinement on droplet break up mechanism for system I and system VII respectively. From Fig. 9(a), it has been noted that the droplet achieves steady state configuration in unbounded domain ($Wc= 0.2$) at $Ca_E=0.6$. But in confined domain ($Wc=0.909$), the deformation characteristic is quite different where the droplet undergoes into a sudden elongation for same value of electric capillary number that ultimately leads the droplet to pinch off into daughter droplets. A similar scenario has also been observed for system VII shown in Fig. 9(b).

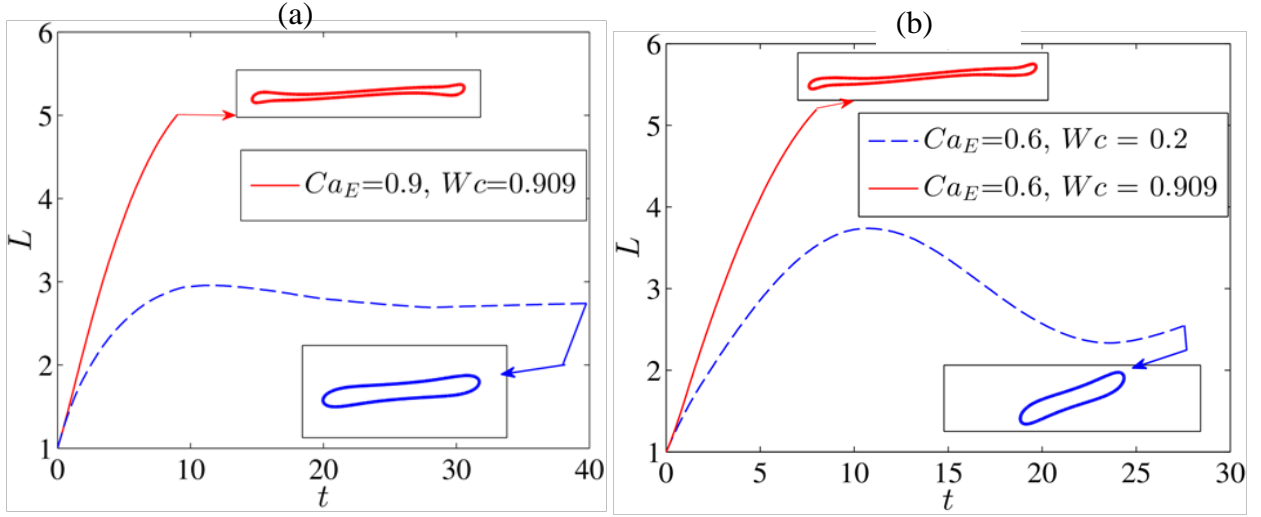

FIG. 9. Effect of confinement on droplet break up for (a). system I having $S=15$ and (b) system VII having $(R, S)=(0.5, 2)$. Other parameters are $Re=0.01$, $\lambda=1$, $M=1$.

A proper physical explanation of the observed behavior is now provided. For perfect dielectric system I, the normal electric stress tries to deform the droplet towards the electrode where as the shear stress tries to retard its effect via deforming the droplet along the flow direction. For system I, the magnitude of normal electric stress is also high as the value of $S$ is very high. Furthermore, domain confinement again strengthens its value that finally leads the droplets to undergoes into breakup phenomenon. The reason behind the deformation phenomenon observed in Fig 9(b) is slightly different. For the system VII, both the normal electric stress and normal hydrodynamic stress tries to deform the droplet into oblate shape in unbounded domain. Likely, the viscous stress also tries to deform the droplet along the flow direction. But, as the effect of domain confinement on the governing stresses is not pronounced, it creates lower deformation and the droplet achieves steady state configuration quickly. But in confined domain, the scenario is quite different. For the considered values of $(R, S)$, the electric field strength in confined domain is very high as shown in Fig 10(a). This necessarily stands for higher normal electric stress (as normal electric stress is proportional to $E^2$ ).Furthermore, the shear rate is also significantly higher that means the magnitude of shear stress is also high in confined domain. It is also worth to note that the reversed hydrodynamic stress tries to reduce the deformation in confined domain. Despite that fact, the droplet undergoes into break up phenomenon due to the combined strength of normal stress and viscous stress.

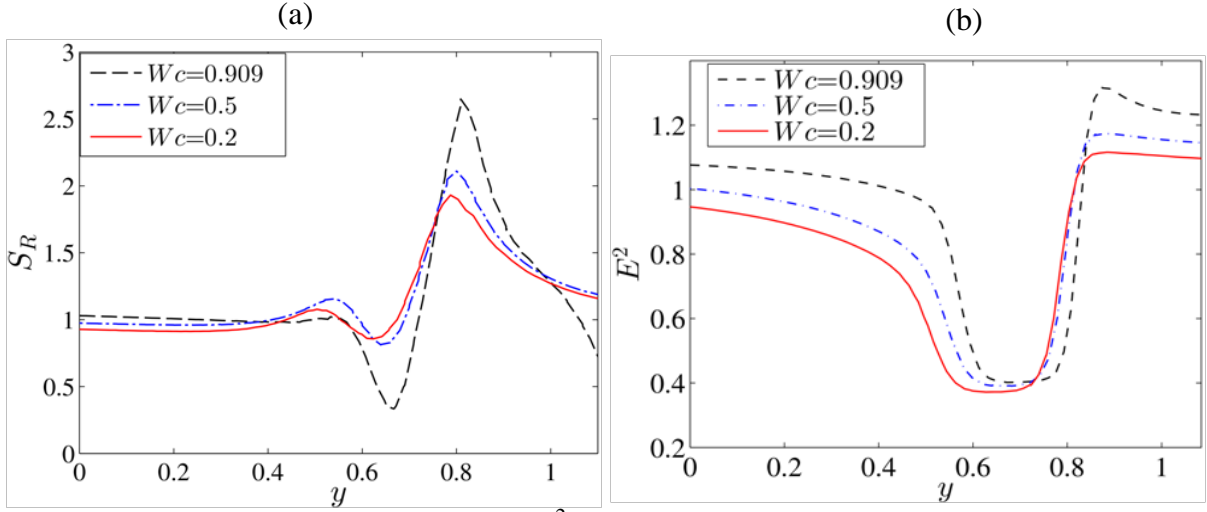

FIG. 10. Variation of (a) shear rate (b) $E^2$ with the vertical distance at the tip of the droplet for system VII having (R, S)=(0.5, 2). Other parameters are $Re$=0.01, $\lambda$ =1, $M$=1.

## VI. CONCLUSIONS

### A. Summary of the findings

In the present analysis, we have investigated the effect of domain confinement on the EHD of droplet under the combine presence of uniform electric field and simple shear flow. For capturing the essential features of the droplet dynamic phenomenon in confined domain, we have performed the numerical simulation in a two dimensional co-ordinate system. In order to validate the numerical result, we have also performed analytical study considering higher order correction term. The present study is focused on the effect of domain confinement on the steady state droplet deformation, transient deformation characteristic and droplet breakup. The key findings are summarized as follow

(i) At lower electric field strength, the perfect dielectric system having $S$<1 shows a monotonically augmentation of the deformation parameter with the confinement ratio. However, non-monotonic variation is observed at higher electric field strength. On the other side, the deformation characteristic follows a non-monotonic variation both at higher and lower electric field strength for perfect dielectric system having $S$>1.

(ii) For both the type of leaky dielectric systems ($S$>$R$ and $R$>$S$), a monotonic increasing behavior of deformation is observed at small electric field strength. But, at higher electric field strength, non-monotonic dependency of deformation parameter on confinement ratio becomes prominent.

(iii) In highly confined domain, for perfect dielectric system having $S$>1, the steady state deformation reduces with electric field strength, whereas the steady state time increases. For perfect dielectric system having $S$<1 also, the value of steady state deformation reduces with electric field strength, but the droplet achieve steady state configuration quickly.

(iv) In highly confined domain, the steady state deformation parameters as well as steady state time increases with electric field strength for leaky dielectric system having $S$>$R$.

On the contrary, both the deformation parameter and steady time decreases with electric field strength for the leaky dielectric system with $R>S$.

(v) For both the perfect dielectric system and leaky dielectric system, the domain confinement incite the droplet breakup phenomenon.

**B. Remarks**

In the present analysis, we have considered a 2D computational domain with drastic simplification. Through Fig 2 and 4, we have shown a good agreement between analytical solution obtained for liquid column and present numerical result and confirmed that the present simulation results are appropriate for capturing the deformation dynamics of liquid column in confined domain in combined presence of background shear flow and transverse electric field. But the same is not correct for liquid droplet as the flows are three dimensional in nature. However, there are some similarities in EHD of 2D and 3D liquid droplet. In presence of uniform electric field, the EHD flow circulation for both the 2D and 3D liquid droplet are similar and consists of four circulation rolls inside and outside of the droplet. For both the 2D and 3D cases, the direction of flow circulation also depends on the electrical property ratios (conductivity ratio and permittivity ratio) of the droplet and suspending fluid. To confirm the applicability of the present 2D model, we have performed a pairs of model comparison test . In first test, we have compared our numerical result with the experimental results (for 3D droplet in unbounded domain) of Salipante and Vlahovska[57] and Tsukada et al.[58] (refer to appendix C for details). The study suggests that the trends of the characteristic curves are very similar and a well agreement is achieved for lower values of $Ca_E$, whereas the deviation is seemed to be higher for higher values of $Ca_E$. In another set of comparison test, we have compared our numerical result with the experimental result of Sibillo et al.[42] (3D droplet suspended in confined shear flow), where also we have found a good matching.

So, after the comparison test, a conclusion can be drawn that the study of 2D droplet dynamics will provide sufficient practical insights into the dynamics of three-dimensional droplet and will uphold substantial degree of physical importance. It is also necessary to note that several literature [36,39,43,45,59,60] have been reported that have considered 2D study for capturing the underlying physics of droplet dynamics.

**APPENDIX A: ANALYTICAL SOLUTION FOR THE DROPLET SHAPE IN UNBOUNDED DOMAIN**

Under the Stokes flow (i.e. $Re=0$) condition, the analytical solution of the present EHD problem can be done by neglecting the effects of walls. In an unbounded domain, at lower value of $Ca$, the strength of the hydrodynamic stress is low that leads to small deformation. Due to that, the droplet shape deviates less from its spherical shape. Again, $M \sim 1$ illustrates that the electric stress is also less that means deformation due to electric field induced electric stress is also less that allows us to obtain a analytical solution of electric potential and velocity field by using regular perturbation method taking $Ca_E$ or $Ca$ as a perturbation parameter. Here, We have considered $Ca$ as a perturbation parameter for sake of convenience. In regular perturbation method, the expansion of each dependent variable takes the following perturbation form [55,61]

$$\chi = \chi^{(0)} + Ca\,\chi^{(Ca)} + Ca^2\,\chi^{(Ca^2)} + O(Ca^3) \qquad (A1)$$

where $\chi^{(0)}$ depicts the leading order term of $\chi$ without shape deformation and $\chi^{(Ca)}$ shows the correction term with the shape deformation.

For a electrostatic problem, the electric potential should satisfy the Laplace equation. The electric potential inside and outside the droplet can be expressed in the following form

$$\phi_i = \sum_{n=1}^{\infty} r^n \sum_{m=1}^{n} \left[ a_{n,m} \cos(m\theta) + \hat{a}_{n,m} \sin(m\theta) \right] \qquad (A2)$$

$$\phi_e = -\mathbf{E}_\infty \cdot \mathbf{r} + \sum_{n=1}^{\infty} \frac{1}{r^n} \sum_{m=1}^{n} \left[ b_{n,m} \cos(m\phi) + \hat{b}_{n,m} \sin(m\phi) \right]. \qquad (A3)$$

The unknown coefficients at each order of perturbation are obtained by applying the appropriate boundary condition at the droplet interface (continuity of electric potential and normal current density). Under the Stokes flow condition, We can expressed the flow field solely in terms of fourth-order bi-harmonic equation for stream function. The general solution of the stream function for the inner and outer phase of the droplet can be expressed in the following form

$$\psi_i = \sum_{n=2}^{\infty} r^n \sum_{m=2}^{n} \left( A_{n,m} \cos(m\theta) + \hat{A}_{n,m} \sin(m\theta) + B_{n,m}(t) r^2 \cos(m\theta) + \hat{B}_{n,m}(t) r^2 \sin(m\theta) \right) \qquad (A4)$$

$$\psi_e = \left(\frac{r^2}{4}\right) - \left(\frac{r^2}{4} \cdot \cos(2\theta)\right) + \sum_{n=2}^{\infty} r^{-n} \sum_{m=2}^{n} \left( \begin{array}{l} C_{n,m} \cos(m\theta) + \hat{C}_{n,m} \sin(m\theta) \\ + E_{n,m} r^2 \cos(m\theta) + \hat{E}_{n,m} r^2 \sin(m\theta) \end{array} \right) \qquad (A5)$$

Here also, we have calculated the unknown coefficient of the Eq.(A4)-(A5) by using the appropriate interfacial boundary condition (no slip and no penetration boundary condition of the velocity and tangential stress boundary condition). By applying the normal stress boundary condition, we have calculated the droplet shape and the important expression of Eq.(17) are represented in the following form

$$L^{(Ca)}{}_{2,0} = -\frac{M(R^2+R-3S+1)}{3(R+1)^2}\left(1-\exp\left(-\frac{t}{\lambda+1}\right)\right)$$

$$\hat{L}^{(Ca)}_{2,0} = \left(1-\exp\left(-\frac{t}{\lambda+1}\right)\right)$$

$$L^{(Ca^2)}{}_{0,1} = -\frac{1}{4}(L_{2,0})^2 - \frac{1}{4}(\hat{L}_{2,0})^2$$

$$L^{(Ca^2)}{}_{2,1} = -\frac{M(R-1)(R^2+R-3S+1)\{l_1+l_2+l_3\}}{72(1+R)^5}\exp\left(-\frac{t}{\lambda+5}\right)$$

$$\hat{L}^{(Ca^2)}_{2,1} = \frac{M(R-1)\{\hat{l}_1+\hat{l}_2+\hat{l}_3\}}{24(R+1)^3}\exp\left(-\frac{t}{\lambda+5}\right)$$

$$L^{(Ca^2)}{}_{4,1} = \frac{\{k_1+k_2+k_3+k_4\}}{720(5\lambda+13)(1+\lambda)(\lambda+2)(1+R^4)}\exp\left(-\frac{6t}{11\lambda+19}\right)$$

$$\hat{L}^{(Ca^2)}_{4,1} = -\frac{M\{\hat{k}_1+\hat{k}_2+\hat{k}_3+\hat{k}_4\}}{360(\lambda+2)(5\lambda+13)(\lambda+1)(R+1)^2}\exp\left(-\frac{6t}{11\lambda+19}\right) \quad (A6)$$

Where

$$l_1 = (R^2\lambda+13R^2-2R\lambda-12S\lambda+38R-28S+\lambda+13)\exp\left(-\frac{4t}{(\lambda+1)(\lambda+5)}\right)$$

$$l_2 = \frac{4(R^2\lambda+13R^2-2R\lambda-12S\lambda+38R-28S+\lambda+13)}{1+\lambda}\exp\left(\frac{t}{(\lambda+5)}\right) \quad (A7)$$

$$l_3 = -\frac{(\lambda+5)(R^2\lambda+13R^2-2R\lambda-12S\lambda+38R-28S+\lambda+13)}{1+\lambda}$$

$$\hat{l}_1 = (R^2\lambda+13R^2-2R\lambda-12S\lambda+38R-28S+\lambda+13)\exp\left(-\frac{4t}{(\lambda+5)(\lambda+1)}\right)$$

$$\hat{l}_2 = \frac{4(R^2\lambda+13R^2-2R\lambda-12S\lambda+38R-28S+\lambda+13)}{\lambda+1}\exp\left(\frac{t}{\lambda+5}\right) \quad (A8)$$

$$\hat{l}_3 = -\frac{(R^2\lambda+13R^2-2R\lambda-12S\lambda+38R-28S+\lambda+13)(\lambda+5)}{\lambda+1}$$

$$k_1 = \begin{pmatrix} 700M^2R^4\lambda^3 + 4400M^2R^4\lambda^2 + 1200M^2R^3\lambda^3 - 4000M^2R^2S\lambda^3 \\ +9068M^2R^4\lambda + 7040M^2R^3\lambda^2 - 24640M^2R^2S\lambda^2 + 1900M^2R^2\lambda^3 \\ -3400M^2RS\lambda^3 + 5700M^2S^2\lambda^3 - 6180R^4\lambda^3 + 6136M^2R^4 \\ +13232M^2R^3\lambda - 49504M^2R^2S\lambda + 11440M^2R^2\lambda^2 - 19360M^2RS\lambda^2 \\ +1200M^2R\lambda^3 + 34320M^2S^2\lambda^2 - 4000M^2S\lambda^3 - 37008R^4\lambda^2 - 24720R^3\lambda^3 \\ +7904M^2R^3 - 32448M^2R^2S + 22300M^2R^2\lambda - 34792M^2RS\lambda + 7040M^2R\lambda^2 \\ +66900M^2S^2\lambda - 24640M^2S\lambda^2 + 700M^2\lambda^3 - 71604R^4\lambda - 148032R^3\lambda^2 - \\ 37080R^2\lambda^3 + 14040M^2R^2 - 19344M^2RS + 13232M^2R\lambda + 42120M^2S^2 \\ -49504M^2S\lambda + 4400M^2\lambda^2 - 44616R^4 - 286416R^3\lambda - 222048R^2\lambda^2 \\ -24720R\lambda^3 + 7904M^2R - 32448M^2S + 9068M^2\lambda - 178464R^3 - 429624R^2\lambda \\ -148032R\lambda^2 - 6180\lambda^3 + 6136M^2 - 267696R^2 - 286416R\lambda - 37008\lambda^2 \\ -178464R - 71604\lambda - 44616 \end{pmatrix} \exp\left(-\frac{6t}{(11\lambda+19)}\right)$$

$$k_2 = \begin{pmatrix} 175M^2R^4\lambda^3 + 965M^2R^4\lambda^2 + 350M^2R^3\lambda^3 - 1050M^2R^2S\lambda^3 \\ +1661M^2R^4\lambda + 1930M^2R^3\lambda^2 - 5790M^2R^2S\lambda^2 + 525M^2R^2\lambda^3 \\ -1050M^2RS\lambda^3 + 1575M^2S^2\lambda^3 - 1575R^4\lambda^3 + 871M^2R^4 \\ +3322M^2R^3\lambda - 9966M^2R^2S\lambda + 2895M^2R^2\lambda^2 - 5790M^2RS\lambda^2 \\ +350M^2R\lambda^3 + 8685M^2S^2\lambda^2 - 1050M^2S\lambda^3 - 8685R^4\lambda^2 \\ -6300R^3\lambda^3 + 1742M^2R^3 - 5226M^2R^2S + 4983M^2R^2\lambda \\ -9966M^2RS\lambda + 1930M^2R\lambda^2 + 14949M^2S^2\lambda - 5790M^2S\lambda^2 \\ +175M^2\lambda^3 - 14949R^4\lambda - 34740R^3\lambda^2 - 9450R^2\lambda^3 + 2613M^2R^2 \\ -5226M^2RS + 3322M^2R\lambda + 7839M^2S^2 - 9966M^2S\lambda \\ +965M^2\lambda^2 - 7839R^4 - 59796R^3\lambda - 52110R^2\lambda^2 - 6300R\lambda^3 \\ +1742M^2R - 5226M^2S + 1661M^2\lambda - 31356R^3 - 89694R^2\lambda \\ -34740R\lambda^2 - 1575\lambda^3 + 871M^2 - 47034R^2 - 59796R\lambda - 8685\lambda^2 \\ -31356R - 14949\lambda - 7839 \end{pmatrix} \exp\left(-\left(\frac{16t(\lambda+2)}{(11\lambda+9)(\lambda+1)}\right)\right)$$

$$k_3 = \begin{pmatrix} 280M^2R^4\lambda^3 + 1184M^2R^4\lambda^2 + 320M^2R^3\lambda^3 - 1440M^2R^2S\lambda^3 \\ +1592M^2R^4\lambda + 640M^2R^3\lambda^2 - 5376M^2R^2S\lambda^2 + 600M^2R^2\lambda^3 \\ -720M^2RS\lambda^3 + 1800M^2S^2\lambda^3 - 2376R^4\lambda^3 + 688M^2R^4 - 320M^2R^3\lambda \\ -6048M^2R^2S\lambda + 1824M^2R^2\lambda^2 - 192M^2RS\lambda^2 + 320M^2R\lambda^3 \\ +5472M^2S^2\lambda^2 - 1440M^2S\lambda^3 - 7776R^4\lambda^2 - 9504R^3\lambda^3 - 640M^2R^3 \\ -2112*M^2R^2S + 1272M^2R^2\lambda + 4464M^2RS\lambda + 640M^2R\lambda^2 \\ +3816M^2S^2\lambda - 5376M^2S\lambda^2 + 280M^2\lambda^3 - 6696R^4\lambda - 31104R^3\lambda^2 \\ -14256R^2\lambda^3 + 48M^2R^2 + 3936M^2RS - 320M^2R\lambda + 144M^2S^2 \\ -6048M^2S\lambda + 1184M^2\lambda^2 - 1296R^4 - 26784R^3\lambda - 46656R^2\lambda^2 \\ -9504R\lambda^3 - 640M^2R - 2112M^2S + 1592M^2\lambda - 5184R^3 \\ -40176R^2\lambda - 31104R\lambda^2 - 2376\lambda^3 + 688M^2 - 7776R^2 \\ -26784R\lambda - 7776\lambda^2 - 5184R - 6696\lambda - 1296 \end{pmatrix} \exp\left(-\left(\frac{t(5\lambda+13)}{(11\lambda+9)(\lambda+1)}\right)\right)$$

$$k_4 = \begin{pmatrix} -1155M^2R^4\lambda^3 - 6549M^2R^4\lambda^2 - 1870M^2R^3\lambda^3 + 6490M^2R^2S\lambda^3 - 12321M^2R^4\lambda \\ -9610M^2R^3\lambda^2 + 35806M^2R^2S\lambda^2 - 3025M^2R^2\lambda^3 + 5170M^2RS\lambda^3 - 9075M^2S^2\lambda^3 \\ +10131R^4\lambda^3 - 7695M^2R^4 - 16234M^2R^3\lambda + 65518M^2R^2S\lambda - 16159M^2R^2\lambda^2 \\ +25342M^2RS\lambda^2 - 1870M^2R\lambda^3 - 48477M^2S^2\lambda^2 + 6490M^2S\lambda^3 + 53469R^4\lambda^2 \\ +40524R^3\lambda^3 - 9006M^2R^3 + 39786M^2R^2S - 28555M^2R^2\lambda + 40294M^2RS\lambda \\ -9610M^2R\lambda^2 - 85665M^2S^2\lambda + 35806M^2S\lambda^2 - 1155M^2\lambda^3 + 93249R^4\lambda + 213876R^3\lambda^2 \\ +60786R^2\lambda^3 - 16701M^2R^2 + 20634M^2RS - 16234M^2R\lambda - 50103M^2S^2 + 65518M^2S\lambda \\ -6549M^2\lambda^2 + 53751R^4 + 372996R^3\lambda + 320814R^2\lambda^2 + 40524R\lambda^3 - 9006M^2R \\ +39786M^2S - 12321M^2\lambda + 215004R^3 + 559494R^2\lambda + 213876R\lambda^2 + 10131\lambda^3 \\ -7695M^2 + 322506R^2 + 372996R\lambda + 53469\lambda^2 + 215004R + 93249\lambda + 53751 \end{pmatrix}$$

(A9)

$$\left.\begin{aligned}
\hat{k}_1 &= \begin{pmatrix} 2080R^2\lambda^3 + 12768R^2\lambda^2 + 1780R\lambda^3 - 5940S\lambda^3 + 25536R^2\lambda \\ +10128R\lambda^2 - 35664S\lambda^2 + 2080\lambda^3 + 16640R^2 + 18180R\lambda \\ -69252S\lambda + 12768\lambda^2 + 10088R - 43368S + 25536\lambda + 16640 \end{pmatrix} \exp\left(\frac{6t}{11\lambda + 9}\right) \\
\hat{k}_2 &= \begin{pmatrix} 525R^2\lambda^3 + 2895R^2\lambda^2 + 525R\lambda^3 - 1575S\lambda^3 + 4983R^2\lambda + 2895R\lambda^2 \\ -8685S\lambda^2 + 525\lambda^3 + 2613R^2 + 4983R\lambda - 14949S\lambda + 2895\lambda^2 + \\ 2613R - 7839S + 4983\lambda + 2613 \end{pmatrix} \exp\left(-\frac{16t(\lambda+2)}{(11\lambda+9)(\lambda+1)}\right) \\
\hat{k}_3 &= \begin{pmatrix} 816R^2\lambda^3 + 3072R^2\lambda^2 + 456R\lambda^3 - 2088S\lambda^3 + 3504R^2\lambda \\ +480R\lambda^2 - 6624S\lambda^2 + 816\lambda^3 + 1248R^2 - 1752R\lambda - 5256S\lambda \\ +3072\lambda^2 - 1776R - 720S + 3504\lambda + 1248 \end{pmatrix} \exp\left(-\frac{t(5\lambda+13)}{(11\lambda+9)(\lambda+1)}\right) \\
\hat{k}_4 &= \begin{pmatrix} -3421R^2\lambda^3 - 18735R^2\lambda^2 - 2761R\lambda^3 + 9603S\lambda^3 - 34023R^2\lambda - \\ 13503R\lambda^2 + 50973S\lambda^2 - 3421\lambda^3 - 20501R^2 - 21411R\lambda + 89457S\lambda \\ -18735\lambda^2 - 10925R + 51927S - 34023\lambda - 20501 \end{pmatrix}
\end{aligned}\right\}$$

(A10)

## APPENDIX B: GRID AND CAHN NUMBER INDEPENDENCE STUDY

Grid independence study and Cahn number independence study have been performed for checking the accuracy of the numerical solution. It is worthy to mention that the grid size and *Cn* number are same near the interface. So, a correct Cahn number independence study must ensure a accurate grid independence study and vice versa. For performing the *Cn* independence

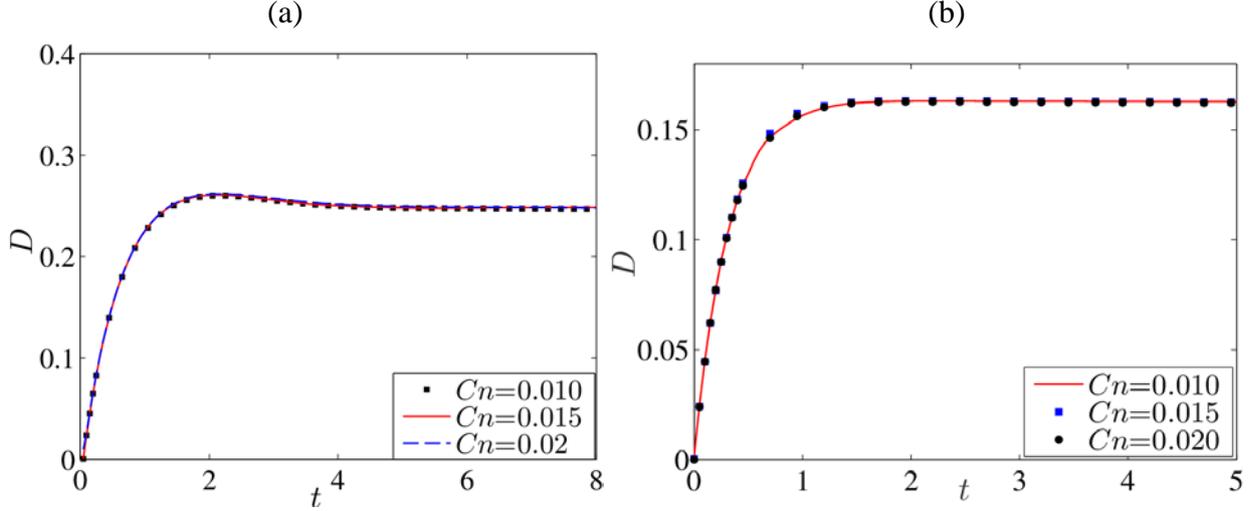

Figure 11. Grid and *Cn* independence study. Deformation of the droplet interface in (a) simple shear flow with $Ca = 0.1$, $Wc=0.8$ and (b) shear flow with electric field having $S = 15$, $Ca = 0.14$, $M = 1$, $Wc = 0.2$

test, deformation parameter (*D*) has been evaluated for two cases : a) the droplet is suspended in a simple shear flow with $Wc = 0.80$, $Ca = 0.1$ and $Re = 0.01$ and b) the droplet is suspended in simple shear flow under a transverse electric field with $Wc =0.20$, $Ca =0.14$, $M =1$ and $Re=0.01$ for three different Cahn numbers (*Cn*=0.02, 0.01, 0.015) as shown in figure. 11(a) and figure. 11(b) respectively. The figure shows a negligible variation in deformation characteristic for the chosen different Cahn numbers. Finally we have chosen *Cn*=0.015 for the present analysis. For the present study, all the plot have been made considering that *Cn*=0.015.

**APPENDIX C: VALIADTION OF THE PRESENT 2D NUMRICAL MODEL**

For conforming the utility of present 2D numerical analysis in to 3D dimensional analysis, we have compared our present numerical result with the experimental result of Sibillo et al.[42] as shown in Fig. 12(a), where a liquid droplet is suspended in a confined shear flow. This study shows a well matching between the numerical result and experimental result. Furthermore ,we have made another comparison of our numerical result with the experimental result of Tsukada et al.[58] & Salipante and Vlahovska[57] as shown in Fig.12(b). The study shows a good matching between the present numerical result and experimental result at lower values of $Ca_E$, where as the deviation is comparatively higher at higher value of $Ca_E$.

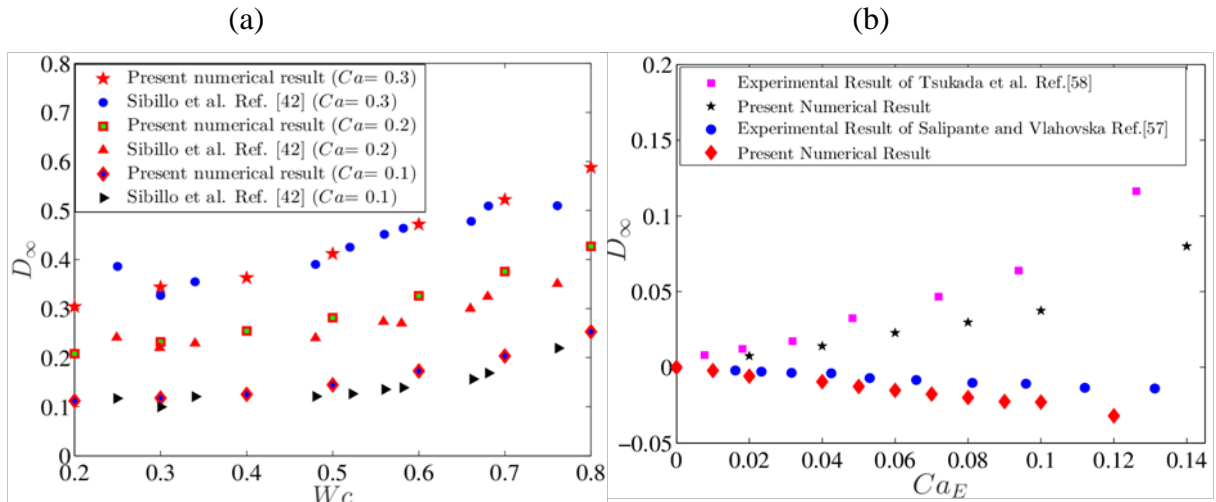

Figure 12. Comparison of present numerical result and experimental result. The properties of fluids considered in the study of Tsukada et al.[58] and Salipante and Vlahovska[57] are ($S$, $R$)=(1.72, 62.4), $\lambda$ =1.47 and ($S$, $R$)=(0.57, 0.027), $\lambda$ =1.40 respectively. The viscosity ratio ($\lambda$) of the fluids used in the study of Sibillo et al.[42] is

**REFERENCES**

[1] H.A. Stone, A.D. Stroock, and A. Ajdari, "ENGINEERING FLOWS IN SMALL DEVICES : Microfluidics Toward a Lab-on-a-Chip," Annu. Rev. Fluid Mech., **36**, 381 (2004).

[2] J.S. Eow and M. Ghadiri, "Drop–drop coalescence in an electric field: the effects of applied electric field and electrode geometry," Colloids Surfaces A Physicochem. Eng. Asp., **219**, 253 (2003).

[3] S. Mhatre, V. Vivacqua, M. Ghadiri, A.M. Abdullah, A. Hassanpour, B. Hewakandamby, B. Azzopardi, and B. Kermani, "Electrostatic phase separation : A review," Chem. Eng. Res. Des., **96**, 177 (2015).

[4] K.J. Ptasinski and P.J. a. M. Kerkhof, "Electric Field Driven Separations: Phenomena and Applications," Sep. Sci. Technol., **27**, 995 (1992).

[5] R. Seemann, M. Brinkmann, T. Pfohl, and S. Herminghaus, "Droplet based microfluidics," Reports Prog. Phys., **75**, 16601 (2012).

[6] Z. Che, T.N. Wong, N.-T. Nguyen, and C. Yang, "Three dimensional features of convective heat transfer in droplet-based microchannel heat sinks," Int. J. Heat Mass Transf., **86**, 455 (2015).

[7] M. Najah, R. Calbrix, I.P. Mahendra-Wijaya, T. Beneyton, A.D. Griffiths, and A. Drevelle, "Droplet-Based Microfluidics Platform for Ultra-High-Throughput Bioprospecting of Cellulolytic Microorganisms," Chem. Biol., **21**, 1722 (2014).

[8] K. Ahn, C. Kerbage, T.P. Hunt, R.M. Westervelt, D.R. Link, and D.A. Weitz, "Dielectrophoretic manipulation of drops for high-speed microfluidic sorting devices," Appl. Phys. Lett., **88**, 24104 (2006).

[9] D.R. Link, E. Grasland-Mongrain, A. Duri, F. Sarrazin, Z. Cheng, G. Cristobal, M. Marquez, and D.A. Weitz, "Electric control of droplets in microfluidic devices.," Angew. Chemie, **45**, 2556 (2006).

[10] J.D. Wehking and R. Kumar, "Droplet actuation in an electrified microfluidic network," Lab Chip, **15**, 793 (2015).

[11] C. T. O'Konski and H.C. ThacherJr, "The Distortion of Aerosol Droplet by an electric field," J. Phys. Chem., **57**, 955 (1953).

[12] G. Taylor, "Studies in Electrohydrodynamics. I. The Circulation Produced in a Drop by Electrical Field," Proc. R. Soc. A, **291**, 159 (1966).

[13] O.O. Ajayi, "A Note on Taylor's Electrohydrodynamic Theory," Proc. R. Soc. A Math. Phys. Eng. Sci., **364**, 499 (1978).

[14] S. Torza R. G. Cox and S. G. Masson, "ELECTROHYDRODYNAMIC DEFORMATION AND BURST OF LIQUID DROPS," Proc. R. Soc. A Math. Phys. Eng. Sci., **269**, 295 (1970).

[15] J.D. Sherwood, "Breakup of fluid droplets in electric and magnetic fields," J. Fluid Mech., **188**, 133 (1988).

[16] J.-W. HA and S.-M. YANG, "Deformation and breakup of Newtonian and non-Newtonian conducting drops in an electric field," J. Fluid Mech., **405**, S0022112099007223 (2000).

[17] E. Lac and G.M. Homsy, "Axisymmetric deformation and stability of a viscous drop in a steady electric field," J. Fluid Mech., **590**, 239 (2007).

[18] R.B. Karyappa, S.D. Deshmukh, and R.M. Thaokar, "Breakup of a conducting drop in a uniform electric field," J. Fluid Mech., **754**, 550 (2014).

[19] S. Mandal, K. Chaudhury, and S. Chakraborty, "Transient dynamics of confined liquid drops in a uniform electric field," Phys. Rev. E, **89**, 53020 (2014).

[20] S. Mandal, A. Bandopadhyay, and S. Chakraborty, "Effect of surface charge convection and shape deformation on the dielectrophoretic motion of a liquid drop," Phys. Rev. E - Stat. Nonlinear, Soft Matter Phys., **93**, 1 (2016).

[21] A.B Basset, "Waves and Jets in a Viscous Liquid Source : , Vol . 16 , No . 1 ( Jan ., 1894 ), pp . 93-110 Published by : The Johns Hopkins University Press Stable URL : http://www.jstor.org/stable/2369834," Am. J. Math., **16**, 93 (1894).

[22] A.L. Huebner and H.N. CHU, "Instability and breakup of charged liquid jets," J. Fluid Mech., **49**, 361 (1971).

[23] P.H. Son and K. Ohba, "Theoritical and experimental investigation on instability of an electrically charged liquid jet," Int. J. Multiph. Flow, **24**, 605 (1998).

[24] C.L. Burcham and D.A. and Saville, "The electrohydrodynamic stability of a liquid bridge: microgravity experiments on a bridge suspended in a dielectric gas," **405**, 37 (2000).

[25] D.. Saville, "Stability of Electrically Charged Viscous Cylinders," Phys. Fluids, **14**, 1095 (1971).

[26] G.I. Taylor, "Electrically driven jet," Proc. R. Soc. A Math. Phys. Eng. Sci., **313**, 453 (1969).

[27] Q. Wang, S. Mählmann, and D.T. Papageorgiou, "Dynamics of liquid jets and threads under the action of radial electric fields: Microthread formation and touchdown singularities," Phys. Fluids, **21**, 32109 (2009).

[28] P.H. Rhodes, R.S. Snyder, and G.O. Roberts, "Electrohydrodynamic Distortion of Sample Streams in Continuous Flow Electrophoresis," J. Colloid Interface Sci, **129**, 78 (1989).

[29] L.L. Limath, L, A. Stoneg, L. Viovy, "Electrohydrodynamic stability of a liquid column under cross fields : Application to continuous flow electrophoresis Electrohydrodynamic stability of a liquid column under cross fields : Application to continuous flow electrophoresis," **10**, 2439 (1998).

[30] D. A. Saville, "Electrohydrodynamic deformation of a particulate stream by a transverse electric field," Phys. Rev. Lett., **71**, 2907 (1993).

[31] D.A. Saville and J.R. Glynn, "Electrohydrodynamic Flows in Nonhomogeneous Liquids," Ind. Eng. Chem. Res., **45**, 6981 (2006).

[32] M. Trau, S. Sankaran, D.A. Saville, and I.A. Aksay, "Pattern Formation in Nonaqueous Colloidal Dispersions via Electrohydrodynamic Flow," **11**, 4665 (1995).

[33] M.N. Reddy and A. Esmaeeli, "The EHD-driven fluid flow and deformation of a liquid jet by a transverse electric field," Int. J. Multiph. Flow, **35**, 1051 (2009).

[34] A. Esmaeeli and P. Sharifi, "The transient dynamics of a liquid column in a uniform transverse electric field of small strength," J. Electrostat., **69**, 504 (2011).

[35] A. Behjatian and A. Esmaeeli, "Electrohydrodynamics of a liquid column under a transverse electric field in confined domains," Int. J. Multiph. Flow, **48**, 71 (2013).

[36] K.S. Sheth and C. Pozrikidis, "Effects of inertia on the deformation of liquid drops in simple shear flow," Comput. Fluids, **24**, 101 (1995).

[37] D.B. Khismatullin, Y. Renardy, and V. Cristini, "Inertia-induced breakup of highly viscous drops subjected to simple shear," Phys. Fluids, **15**, 1351 (2003).

[38] M.R. Kennedy, C. Pozrikidis, and R. Skalak, "Motion and deformation of liquid drops, and the rheology of dilute emulsions in simple shear flow," Comput. Fluids, **23**, 251 (1994).

[39] H. Zhou and C. Pozrikidis, "The flow of suspensions in channels: Single files of drops," Phys. Fluids A Fluid Dyn., **5**, 311 (1993).

[40] N. Ioannou, H. Liu, and Y.H. Zhang, "Droplet dynamics in confinement," J. Comput. Sci., **17**, 463 (2016).

[41] A. Vananroye, P. Van Puyvelde, and P. Moldenaers, "Deformation and orientation of single droplets during shear flow: combined effects of confinement and compatibilization," Rheol. Acta, **50**, 231 (2011).

[42] V. Sibillo, G. Pasquariello, M. Simeone, V. Cristini, and S. Guido, "Drop Deformation in Microconfined Shear Flow," Phys. Rev. Lett., **97**, 1 (2006).

[43] N. Barai and N. Mandal, "Breakup modes of fluid drops in confined shear flows," Phys. Fluids, **28**, 73302 (2016).

[44] R.S. Allan and S.G. Mason, "Particle Behaviour in Shear and Electric Fields. I. Deformation and Burst of Fluid Drops," Proc. R. Soc. A Math. Phys. Eng. Sci., **267**, 45 (1962).

[45] Mahlmann S. and Papageorgiou D.T, "Numerical study of electric field effects on the deformation of two-dimensional liquid drops in simple shear flow at arbitrary Reynolds number ¨," J. Fluid Mech., **626**, 367 (2009).

[46] VLAHOVSKA. P. M., "On the rheology of a dilute emulsion in a uniform electric field," J. Fluid Mech., **670**, 481 (2011).

[47] S. Mandal, S. Sinha, A. Bandopadhyay, and S. Chakraborty, "Drop deformation and emulsion rheology under the combined influence of uniform electric field and linear flow," J. Fluid Mech., **841**, 408 (2018).

[48] D.A. Saville, "Electrohydrodynamics:The Taylor-Melcher Leaky Dielectric Model," Annu. Rev. Fluid Mech., **29**, 27 (1997).

[49] S. Mandal, A. Bandopadhyay, and S. Chakraborty, "Effect of interfacial slip on the cross-stream migration of a drop in an unbounded Poiseuille flow," Phys. Rev. E, **92**, 23002 (2015).

[50] D. Jacqmin, "Calculation of two-phase Navier–Stokes flows using phase-field modeling," J. Comput. Phys., **155**, 96 (1999).

[51] V.E. Badalassi, H.D. Ceniceros, and S. Banerjee, "Computation of multiphase systems with phase field models," J. Comput. Phys., **190**, 371 (2003).

[52] S. Mandal, U. Ghosh, A. Bandopadhyay, and S. Chakraborty, "Electro-osmosis of superimposed fluids in the presence of modulated charged surfaces in narrow confinements," J. Fluid Mech., **776**, 390 (2015).

[53] K. Chaudhury, S. Mandal, and S. Chakraborty, "Droplet migration characteristics in confined oscillatory microflows," Phys. Rev. E, **93**, 1 (2016).

[54] X.-P. Wang, T. Qian, and P. Sheng, "Moving contact line on chemically patterned surfaces," J. Fluid Mech., **605**, 59 (2008).

[55] X. Xu and G.M. Homsy, "The settling velocity and shape distortion of drops in a uniform electric field," J. Fluid Mech., **564**, 395 (2006).

[56] A. Bandopadhyay, S. Mandal, and N.K. Kishore, "Uniform electric-field-induced lateral migration of a sedimenting drop," J. Fluid Mech., **792**, 553 (2017).

[57] P.F. Salipante and P.M. Vlahovska, "Electrohydrodynamics of drops in strong uniform dc electric fields," Phys. Fluids, **22**, 112110 (2010).

[58] T. Tsukada, T. Katayama, Y. Ito, and M. Hozawa, "Theoretical and Experimental Studies of Circulations Inside and Outside a Deformed Drop under a Uniform Electric Field.," J. Chem. Eng. JAPAN, **26**, 698 (1993).

[59] S. Mortazavi and G. Tryggvason, "A numerical study of the motion of drops in Poiseuille flow. Part 1. Lateral migration of one drop," J. Fluid Mech., **411**, S0022112099008204 (2000).

[60] C.A. Stan, L. Guglielmini, A.K. Ellerbee, D. Caviezel, H.A. Stone, and G.M. Whitesides, "Sheathless hydrodynamic positioning of buoyant drops and bubbles inside microchannels," Phys. Rev. E, **84**, 36302 (2011).

[61] A. Bandopadhyay, S. Mandal, N.K. Kishore, and S. Chakraborty, "Uniform electric-field-induced lateral migration of a sedimenting drop," J. Fluid Mech., **792**, 553 (2016).